\documentclass[a4paper,12pt]{article}
\usepackage{jheppub}
\usepackage{amsmath}
\usepackage{amsthm}
\usepackage{amsfonts}
\usepackage{amssymb}
\usepackage{hyperref}
\usepackage{textcomp}
\usepackage{graphicx}
\usepackage{bm}
\usepackage{color}
\usepackage{verbatim}

\usepackage{graphicx,color}
\usepackage{epstopdf}
\usepackage{epsfig}

\usepackage{amsfonts,amsmath,amssymb}

\newcommand{\p}{\partial}

\newcommand{\rar}{\rightarrow}

\begin{document}

 \title{Holographic discommensurations}
  
 \author[a,b]{Alexander Krikun\footnote{https://orcid.org/0000-0001-8789-8703}}

\affiliation[a]{Institute-Lorentz for Theoretical Physics, $\Delta ITP$,
Leiden University \\ PO Box 9506, Leiden 2300 RA, The Netherlands}

\affiliation[b]{Institute for Theoretical and Experimental Physics (ITEP)\footnote{On leave from}, \\ B. Cheryomushkinskaya 25, 117218 Moscow, Russia }


\emailAdd{krikun@lorentz.leidenuniv.nl}

\abstract{When the system with internal tendency to a spontaneous formation of a spatially periodic state is brought in contact with the external explicit periodic potential, the interesting phenomenon of commensurate lock in can be observed. In case when the explicit potential is strong enough and its period is close to the period of the spontaneous structure, the latter is forced to assume the periodicity of the former and the commensurate state becomes a thermodynamically preferred one. If instead the two periods are significantly different, the incommensurate state is formed. It is characterized by a finite density of solitonic objects -- discommensurations -- on top of the commensurate state. In this note I study the properties of discommensurations in holographic model with inhomogeneous translational symmetry breaking and explain how one can understand the commensurate/incommensurate phase transition as a proliferation of these solitons. Some useful numerical techniques are discussed in the Appendix.}

\maketitle

\section{Introduction}

Different kinds of spatially modulated structures which break translation symmetry either explicitly or spontaneously are abundant in condensed matter systems. It all starts from the crystal lattice, which forms a basis for any theoretical description of the condensed matter and breaks translations and rotations down to the discrete groups associated with Bloch momentum and crystal symmetry. On top of that some most interesting systems, including high temperature superconductors, demonstrate the spontaneous formation of the superstructures in the form of charge and spin density waves. Clearly, the interplay between these explicit and spontaneous mechanisms of translation symmetry breaking is interesting. 

In holographic models of condensed matter systems (AdS/CMT) the status of translation symmetry breaking is different. This is an additional ingredient which one has to introduce on top of the Lorentz invariant theory of gravity. Firstly, the spontaneous translation symmetry breaking has been considered in \cite{Ooguri:2010kt, Donos:2012wi,Donos:2013wia,Donos:2013gda,Withers:2013loa,Withers:2014sja} and later on the explicit potentials have also been introduced \cite{Horowitz:2012gs,Horowitz:2012ky,Donos:2012js,Donos:2013eha,Andrade:2013gsa}. Only recently the interplay between these different mechanisms attracted some attention in the context of pinning of the spontaneous superstructure, pseudo-Goldstone modes and phonons \cite{Delacretaz:2016ivq,Delacretaz:2017zxd,Andrade:2017cnc,Amoretti:2016bxs,Jokela:2016xuy,Jokela:2017ltu,Alberte:2017cch}. But more importantly for the present study, the \textit{commensurate lock in} between explicit and spontaneous structures have been studied in \cite{Andrade:2015iyf,Andrade:2017leb,Mott}. In \cite{Mott} it has been shown that the commensurate state in holographic model provides a description to the Mott insulator, and the doping can be understood as departure from the commensurate to incommensurate or higher order commensurate states. In this note I will study in detail the mechanism behind commensurate/incommensurate phase transition in the holographic setup of \cite{Mott}.

The paper is organized as follows. In the rest of the Introduction we'll discuss a toy model, which will help to set up the necessary notions and intuition. In Sec.\ref{sec:the_holographic_setup} the holographic setup will be introduced and the commensurate state will be discussed. Sec.\ref{sec:nonlin_solution} is devoted to the discussion of the peculiarities in obtaining the nonlinear incommensurate solutions. In Sec.\ref{sec:discom} I will study the features of the discommensurations and address their role in commensurate/incommensurate phase transition. Two Appendices describe the numerical techniques and precision control.

\begin{figure}[ht]
\includegraphics[width=0.49\linewidth]{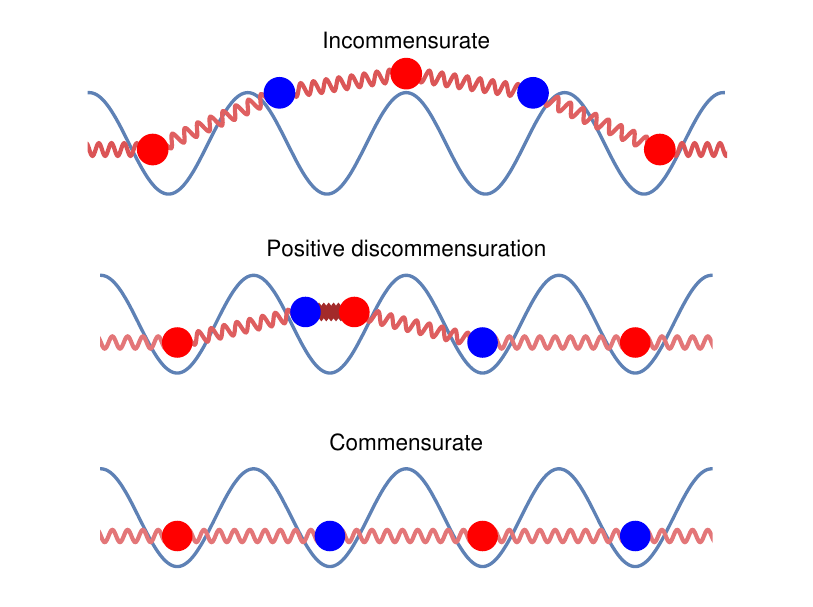}
\includegraphics[width=0.49\linewidth]{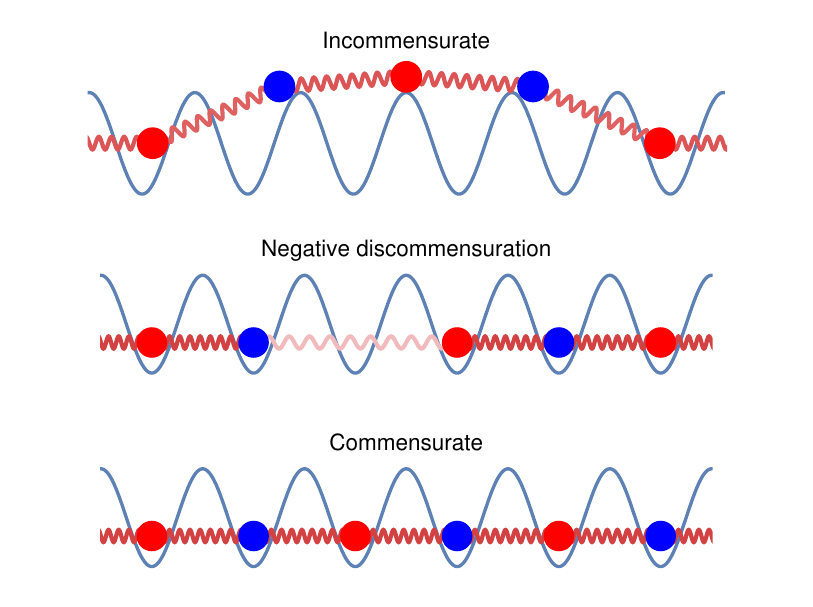}
\caption{\label{Fig:disc_cart}\textbf{Cartoon of discommensurations.} Left panel: $\lambda_k>\lambda_{p_0}$, there are more particles then potential minima, positive discommensuration is formed. Right panel: $\lambda_k<\lambda_{p_0}$, fewer particles then minima, negative discommensuration is formed. Note in both cases the discommensuration changes the staggered ``red-blue'' order in comparison with the ``parent'' commensurate state.}
\end{figure}

The effect of commensurate lock in between two periodic structures is well known in physics \cite{bak1982commensurate,mcmillan1976theory,pokrovsky1979ground,FKbook}. The simplest system which demonstrates it is the Frenkel-Kontorova model, which considers a set of point particles, connected with springs, lying on top of the periodic lattice potential. In case when the potential is absent, the springs assume the normal state and the particles form a periodic structure with spontaneous period $\lambda_{p_0}$. Clearly if one turns on the potential with exactly the same period $\lambda_k = \lambda_{p_0}$, the particles will just fall in the minima of the potential and the springs will not be deformed. The state where the resulting periods of the explicit and spontaneous structures are equal $\frac{\lambda_k}{\lambda_p} = 1 $ is called the \textit{lowest order commensurate}. What will happen if the potential has a different period $\lambda_k > \lambda_{p_0}$? Then there are several possibilities. If the potential is strong enough, then the springs will be forced to stretch, in order that all particles fall into minima, and assume the modified period $\lambda_p = \lambda_k$ (Fig.\ref{Fig:disc_cart}, bottom left). This state is commensurate and the spontaneous structure is called \textit{commensurately locked in} by the lattice. If instead the springs are very strong, and potential relatively weak, then some particles will acquire the additional potential energy, keeping the springs from stretching (Fig. \ref{Fig:disc_cart}, top left). This is \textit{incommensurate} state since the resulting period of the spontaneous structure has no relation to the lattice spacing $\lambda_p \neq \lambda_k$. This incommensurate state is characterized by the feature that at any large enough sample of the system there are more particles then the minima of the potential. 

If the potential in the incommensurate state becomes stronger, more and more particles fall in the minima, locally deforming the springs. Eventually the mismatch in the number of particles gets localized in the single unit cell of the lattice (Fig.\ref{Fig:disc_cart}, middle left). The resulting state looks like the commensurate state everywhere, except around a local defect where two particles fall in the same potential well. This defect is a \textit{discommensuration} -- the soliton on top of the commensurate state, which accounts for the mismatch of the periods of two structures. Clearly, the discommensuration can be defined only if the state is not too far from commensurate point: ${\lambda_p}/{\lambda_k} - 1 \ll 1$,\footnote{Strictly speaking, discommensurations can be considered around the other, higher order commensurate points too, i.e. ${\lambda_p}/{\lambda_k}=2, \dots$, but we focus on the leading commensurate case here.} otherwise the density of them gets so high that one can not define a ``parent'' state. If the particles are charged, then the discommensuration, having one excess particle in the unit cell, bears positive charge. In complete analogy one can consider the \textit{negative} discommensuration, which arises when $\lambda_k < \lambda_{p_0}$ and there are fewer particles then the potential wells (Fig.\ref{Fig:disc_cart}, right). This one is seen as an empty potential well and bears a unit negative charge. Interestingly, if we now introduce the additional $\mathbb{Z}^2$ quantum number and assume that the dynamical system tends to form ``anti-ferromagnetic'' order, then both types of discommensurations will act as domain walls in this staggered order.

In what follows I will study the similar features of discommensurations, which arise in the holographic model with spontaneous and explicit periodic structures.

\section{The holographic setup} 
\label{sec:the_holographic_setup}

Consider the model of \cite{Donos:2011bh}, used in \cite{Mott}. It includes dynamical gravity with negative cosmological constant in 3+1 dimensions, $U(1)$ gauge field and a pseudoscalar field, axion, which is coupled to the $\theta$-term. The action reads \cite{Withers:2013kva, Withers:2013loa}:
\begin{equation}\label{S_0}
	S = \int d^4 x \sqrt{- g} \left( R - \frac{1}{2} (\partial \psi)^2 - \frac{\tau(\psi)}{4} F^2 - V(\psi) \right)
	 - \frac{1}{2} \int {\vartheta}(\psi) F \wedge F
\end{equation}
Here $F = d{\cal A}$ is the field strength of the $U(1)$ gauge field ${\cal A}$. Following 
\cite{Mott,Donos:2011bh, Withers:2013kva, Withers:2013loa, Donos:2013wia}, the couplings are chosen to be 
\begin{gather}
\label{equ:potentials}
	V(\psi) \equiv 2 \Lambda +W(\psi) = - 6 \cosh (\psi /\sqrt{3}) , \quad \\
	\notag
	 \tau(\psi) = {\rm sech} (\sqrt{3} \psi), \quad \vartheta(\psi) = \frac{c_1}{6 \sqrt{2}} \tanh(\sqrt{3} \psi),
\end{gather}
Note that in these conventions the cosmological constant is $\Lambda = - 3$ and the mass of the scalar is $m^2 = -2$. 

It was shown in \cite{Donos:2011bh, Withers:2013kva, Withers:2013loa, Donos:2013wia} that due to the $\vartheta$-coupling this model develops an instability at low temperature evolving into the spatially modulated ground state, which breaks translations spontaneously and has a periodic charge density (CDW) with wavelength $\lambda_{p_0} = 2 \pi / \mathbf{p}_0$ \footnote{Here, following \cite{Mott} we define the wavelength with respect to the charge density modulation, which is one half of the wavelength of the oscillating current.}, where $\mathbf{p}_0(T)$ is a temperature dependent thermodynamically preferred momentum. This is in complete analogy to the system with springs discussed in the Introduction. Noteworthy, this inhomogeneous state features the oscillating diamagnetic currents $J_y \sim \cos(x \mathbf{p}_0/2)$ on the boundary. The features of this state depend on the value of the coupling $c_1$ in \eqref{equ:potentials} and we'll focus on $c_1=17$ used in \cite{Mott}.

Furthermore, one can introduce the background lattice potential, which would break translations explicitly. Following \cite{Horowitz:2012ky,Donos:2014yya,Rangamani:2015hka} it can be done by 
turning on a spatially modulated chemical potential, $A_t(z=0,x) = \mu(x)$, with 
\begin{equation}\label{eq:mu x}
	\mu(x) = \mu_0 (1 + A \cos(\boldsymbol{\mathit{k}} x) ).
\end{equation}
Featuring two spatial modulation scales: $\mathbf{p}_0$ and $\mathbf{k}$, this setup is sufficient to address the interesting physics of commensurability. 

At large temperature in absence of the explicit potential the ground state of \eqref{S_0} is the translational invariant Reissner-Nordstr\"om (RN) black hole 
\begin{equation}\label{RN soln}
	ds^2 = \frac{1}{z^2} \left( - f(z) dt^2 + \frac{dz^2}{f(z)} + dx^2 + dy^2 \right) , \quad A = \bar \mu (1 - z) dt , \qquad \psi = 0
\end{equation}
\noindent where 
\begin{equation}\label{f RN}
	f = (1-z)\left( 1 + z + z^2 - \bar \mu^2 z^3 /4 \right)
\end{equation}
with temperature
\begin{equation}\label{eq:T}
 	\boldsymbol{\mathit{T}} = \frac{12 - \bar \mu^2}{16 \pi}.
\end{equation} 
The conformal boundary is located at $z=0$ while the black hole horizon is at $z=1$.
Without loss of generality, 
one can set $\mu_0 = \bar \mu$. We will express the dimensionful parameters of the model, denoted up until now in bold script, in units of $\bar \mu$ by making the replacements
\begin{equation}
\label{mu units}	
		\boldsymbol{\mathit{T}} = T \bar \mu, \qquad  \boldsymbol{\mathit{k}} = k \bar \mu, \qquad 
		\boldsymbol{\mathit{p}} = p \bar \mu
\end{equation}

It can be shown that the ansatz
\begin{align}\label{ds2 anstaz}
	ds^2 &= \! \frac{1}{z^2}\left( \! - Q_{tt} f(z) dt^2 + Q_{zz} \frac{dz^2}{f(z)} + Q_{xx} (dx + Q_{zx} dz)^2 + Q_{yy} ( dy + Q_{ty} dt )^2  \right), \\
	{\cal A} &= A_t dt + A_y dy 
\end{align}
with all unknown functions dependent on the holographic coordinate $z$ and the boundary coordinate $x$, is sufficient to obtain the spatially modulated solutions of interest. 

Following the standard AdS/CFT prescription, the boundary values of the holographic fields are dual to the one-point functions in the boundary theory:
\begin{align}
\label{series UV 1}
	Q_{tt} & = 1 + z^2 Q_{tt}^{(2)}(x) + z^3 Q_{tt}^{(3)}(x) + O(z^4) \\
  A_t &= \mu(x) - z \rho(x) + O(z^2) \\
  A_y & = z J_y(x) + O(z^2), \\
  \epsilon(x) &= 2 + \frac{\bar \mu^2}{2} - 3 Q_{tt}^{(3)}(x)  
\end{align}
where $\mu$ is a spatially modulated chemical potential, $\rho$ is a charge density, $J_y$ -- diamagnetic current and $\epsilon$ -- the energy density. 

We look for finite temperature solutions, therefore near horizon all functions must be regular. In addition, the equations of motion require $Q_{tt}(1,x) = Q_{zz}(1,x)$, which in turn implies that the surface gravity is constant and given by \eqref{eq:T}, see e.g.\cite{Horowitz:2012ky}.

It is relatively easy to construct the commensurate state. One starts from the spontaneous solution at certain finite temperature $T$ (in this note I will consider fixed $T=0.01$) with thermodynamically preferred momentum $p=p_0(T)$ and then turn on the explicit lattice with \textbf{exact} same momentum $k=p$. Then the two structures will be commensurate by construction. One can use the standard DeTurck method \cite{Headrick:2009pv, Wiseman:2011by} in the periodic computation domain with period $\lambda = 4\pi/k$\footnote{Note here once again that we are working with the momenta of the charge modulation, which are twice larger then the momenta of the current/axion modulation. Therefore one actually needs to consider the computation domain of twice the period of the CDW in order to accommodate one period of the current.}. As it has been shown in detail in \cite{Mott}, this state is the holographic analogue of the Mott insulator. This will be the ``parent''  commensurate state, mentioned in the Introduction. The features of this $1/1$ state are mostly similar to those of the pure spontaneous crystal: the staggered diamagnetic currents are seen and the charge modulation is present as well \cite{Mott}.

The focus of this note is on the thermodynamic stability of this state and its transition to the incommensurate one. But in order to study thermodynamic stability, we need to construct the competing incommensurate solutions first and compare the spatially averaged thermodynamic potential, given by
\begin{equation}\label{eq:free energy}
	 \Omega(x) = \epsilon(x) - \boldsymbol{\mathit{T}} s(x) - \mu(x) \rho(x)
\end{equation}
where $s$ is the entropy density.

\section{Full backreacted solutions} 
\label{sec:nonlin_solution}

In order to study the thermodynamically preferred phases one has to construct fully backreacted nonlinear solutions corresponding to the coexisting ionic lattice and spontaneous crystal structures. There is a peculiar technical difficulty, which arises as soon as one addresses the nonlinear solutions. In \cite{Andrade:2017leb} the spontaneous striped instability was considered in the perturbative regime. At the linear order we were able to introduce the continuous "Bloch momentum", characterizing the spontaneous structure which had not be proportional to the lattice period in any way. In case of finite amplitude of the striped structure this can not be done anymore. The technical reason is that the $e^{i p x}$ multipliers can not be factored out from the nonlinear equations of motion and one has to rely on the position space representation, where the spontaneous structure is characterized by a certain period $\lambda_p$, which is not necessarily equal to the period of the ionic lattice $\lambda_k$. 

\begin{figure}[ht]
\centering
\includegraphics[width=0.8 \linewidth]{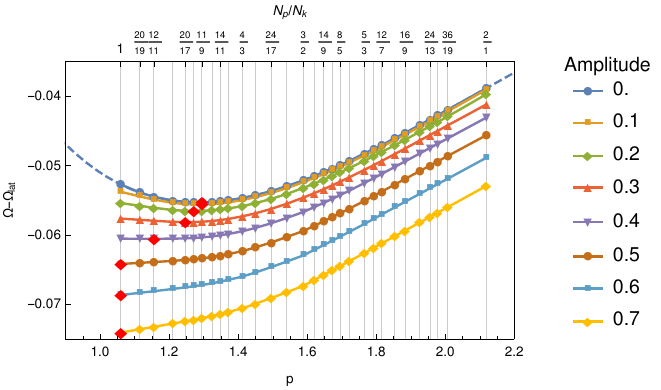}
\includegraphics[width=0.8 \linewidth]{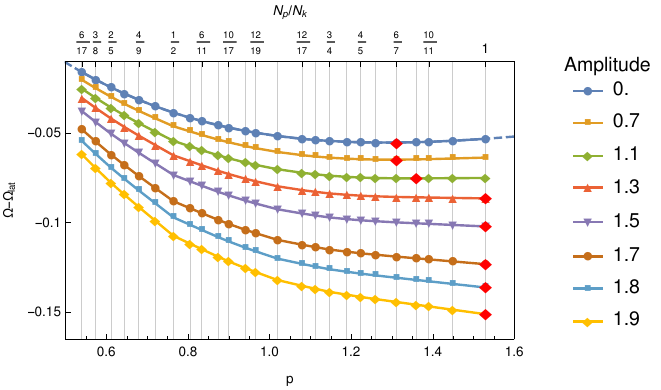}
\caption{\label{fig:wcurves} \textbf{Thermodynamic potential} of different incommensurate solutions, measured with respect to the lattice without spontaneous structure at different amplitudes of the lattice. For temperature $T=0.01$ the spontaneous momentum is $p_0(T) = 1.28$. Upper panel shows the case when the lattice momentum is smaller: $k =1.06<p_0$, Lower panel -- larger $k =1.53>p_0$. Red diamonds show the thermodynamically stable solutions. Both cases demonstrate incommensurate to commensurate phase transition when the lattice gets stronger.}
\end{figure}

In order to set up the numerical Partial Differential Equation (PDE) solver procedure one has to specify only one scale corresponding to the the size of the computation domain with periodic boundary conditions. At this point we see, that in practice one can only access the values $\lambda_p$ which are the rational multiples of $\lambda_k$:
\begin{equation}
\label{eq:rational}
 \lambda_p = \frac{N_k}{N_p} \lambda_k, \qquad N_k, N_p \in \mathbb{N}. 
 \end{equation}
In this case one can choose the computation domain of the size $N_k \lambda_k$ equal to the integer number of lattice periods, which would simultaneously accommodate $N_p$ periods of stripes.\footnote{This situation is completely analogous to the "magnetic unit cell" phenomenon, which arises when one considers a crystal in external magnetic field. The unit cell in this case must simultaneously accommodate integer number of the crystal plaquetes and magnetic fluxes, and can become substantially large. \cite{lifshitz2013statistical}} We see here that the accessible range of spontaneous structure wave-vectors $p$ is now discrete and the density of them is limited by the maximal size of the computation domain, which we can handle in our numerical analysis. In what follows I will use the computation domains including up to $N_k =20$ periods of the lattice or up to $N_p = 40$ periods of the spontaneous CDW\footnote{Note once again, that this corresponds to 20 periods in the diamagnetic currents}, which allows to achieve reasonably dense mesh in our plots for the thermodynamic potential of these solutions (see Fig.\ref{fig:wcurves}).

One might object that this technicality immediately renders it impossible to access the \textit{true} incommensurate solutions in the mathematical sense, where $\lambda_p/\lambda_k$ must be \textit{irrational number}. But I would stress that due to the density of the rational numbers, in physics there is no way to distinguish between high order rational and irrational number. The practical definition of the commensurate (and higher order commensurate) state in this case will be the state where both numbers $N_p$ and $N_k$ are \textit{small} integers. To certain extend in reality the maximum value of $N_k$ is restricted by the quality of the crystal, i.e. the size of the patch of a crystal lattice without defects, or the correlation length of the spatial order. In this regard having $N_k \approx 20$ gives us a reasonable approximation to \textit{physically} incommensurate numbers.   

At this point it is worth mentioning that because the striped structure is spontaneous, unlike the externally sourced lattice, one does not have a direct control over it's shape, including the number of periods and the relative phase (shift) between the stripe and the lattice. In principle the energetically preferred values are set by the dynamics and should be achieved by following the trajectories prescribed by the equations of motion. This is true for the relative shift, which is a continuous parameter. We observe that if we start computation with the different seeds corresponding to different alignments of the stripe and lattice structures, the numerical procedure always converges to the same solution, corresponding to the preferred value of the relative shift. But due to the above mentioned effect, the set of accessible values of the wave-vector is discrete and the system can not smoothly propagate between them. By choosing the initial seed solution with a given periodicity, regarding the numerical relaxation procedure (see Appendix\,\ref{sec:numerical_techniques}) as adiabatic process, we ensure that the system will converge to the state with prescribed number of the spontaneous periods. This state though is only a local minimum of the thermodynamic potential and may be a false vacuum, so one has to construct the solutions with all possible numbers of spontaneous periods and compare their thermodynamic potentials in order to find the true ground state. See Fig.\ref{fig:wcurves}.

Practically, in order to construct the solution with $N_p$ CDW periods on top of the $N_k$ lattice cells with period $\lambda_k$ and amplitude $A$ we first find the spontaneous stripe solution with specific period $\lambda_p$ from \eqref{eq:rational} on top of the translationary symmetric background. Then we catenate $N_p$ copies of these stripes fitting them in the enlarged calculation domain. At this point we turn on $N_k$ periods of the background lattice by slowly changing the boundary condition for the chemical potential, eventually achieving the desired value of $A$ in \eqref{eq:mu x}. This adiabatic process preserves the initial number of the CDW periods what we check numerically at every stage by counting the number of zeros of the oscillating $A_y$ field at the horizon (see Fig.\ref{fig:BulkProfs}).

In complete analogy with the perturbative study of \cite{Andrade:2017leb}, in order to explore the phase diagram at given temperature we first choose the period and the amplitude of the explicit lattice. Then we construct a set of nonlinear solutions, corresponding to the spontaneous structures with different wave-vectors $\lambda_p$ on top of this lattice. We calculate the thermodynamic potential \eqref{eq:free energy} for these solutions $\Omega(\lambda_k)$ and we find the one which is thermodynamically preferred. The sample of the $\Omega(\lambda_k)$ curves, which we get, is shown on Fig.\,\ref{fig:wcurves}. As an example throughout this note I will use two values of the lattice momentum: $k=1.06$ and $k=1.53$. At the temperature under consideration $T=0.01$ the spontaneous momentum of the CDW is $p_0(T) = 1.28$. Therefore the former lattice has longer wavelength then the CDW, promoting positive discommensurations and the latter has shorter wavelength, promoting negative discommensurations.

There are several features on Fig.\,\ref{fig:wcurves}, worth noticing. Firstly, at $A=0$ the  solutions follow the curve which one would obtain for the spontaneous striped solutions on the homogeneous RN background \cite{Withers:2013loa,Withers:2013kva,Donos:2013wia}. I've checked that for $c_1=9.9$ the results coincide with Fig.\,2 in \cite{Withers:2013kva}, which is a valuable check of the numerical method. 

As the amplitude of the lattice is increased, the $\Omega(\lambda_k)$ curves start to deviate smoothly from the RN case. Interestingly, on both of the plots ``one half'' of the curve at finite $A$ is missing. In case $k <  p_0$ it is the part of the curve with $2 \pi \lambda_p^{-1} < k$, in case $k> p_0$ -- the one with $2 \pi \lambda_p^{-1} > k$. The technical reason is that in these cases as we turn on finite amplitude of the chemical potential for a prescribed number of stripes in the seed solution, the numerical procedure does not converge, or converges to the solution with different number of stripes. More precisely, if for $k=1.06$ (upper panel) we choose a seed with 10 CDWs and turn on 11 periods of the lattice (this would correspond to the point on the "missing" left shoulder $N_p/N_k = 10/11 < 1$ on the plot), we will see that the resulting solution has 12 CDWs (now sitting on the right shoulder  $N_p/N_k = 12/11 > 1$). Effectively, one period of the CDW is created dynamically and we can not construct the desired solution with $N_p=10$. In principle this kind of behavior is allowed since the stripes are completely spontaneous. The phenomenon is quite robust: we observe the disappearance of one half of the curve everywhere in the parameter space, which we study. I will address the physical reason behind it in the following Section. 

As one can see, even though we have access only to the discrete set of values, they lie on the smooth curves which have well defined minima. There are two distinct possibilities: the minimum thermodynamic potential is achieved for the period of the stripe close to its spontaneous value $\lambda_{p_0}$, or for the period which is proportional to the lattice spacing\footnote{For the parameters region which we are considering the leading commensurate point is order 1, corresponding to $N_p/N_k = 1$}. The former possibility defines the incommensurate state, the latter -- commensurate lock in. One can see that as the amplitude rises the minimum smoothly shifts from the incommensurate to commensurate point. Henceforth by rising the amplitude we observe the smooth, at least second order phase transition. 

\section{\label{sec:discom} Discommensurations}

\begin{figure}[ht]
\center
\includegraphics[width=0.9 \linewidth]{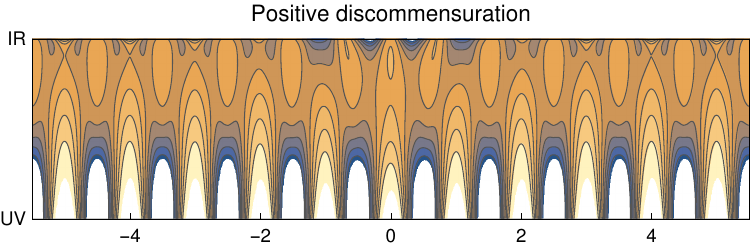}
\includegraphics[width=0.9 \linewidth]{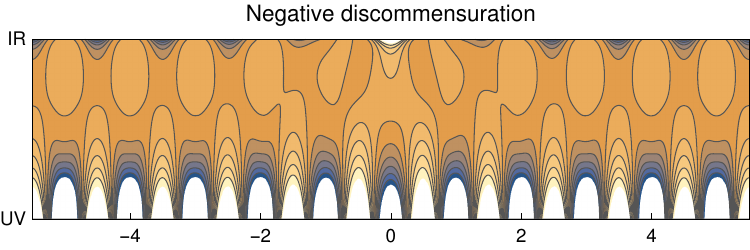}
\caption{\label{fig:BulkProfs} \textbf{Bulk profile of a discommensuration}. Shown is the density of electric field in the bulk $\p_z A_0$, corresponding to the charge density on UV boundary (bottom). The excess/lack of one period of the spontaneous structure is clearly seen near horizon (top). Outside the core of a disc. the profile of the solution is identical to that of the commensurate state.}
\end{figure}

Let's now consider the incommensurate state. As mentioned earlier, the numerical computation in this case is technically more involved, as the numbers of periods in \eqref{eq:rational} can become large.  It is instructive to start the discussion by focusing on the solution which is closest to commensurate $\frac{N_p}{N_k} ={} \frac{1}{1}$ value. Given that
\begin{equation}
\label{discom_counting}
\left| \frac{N_p}{N_k} - \frac{1}{1} \right|= \frac{|N_p - N_k|}{N_k},  
\end{equation}
I will choose $N_k = N_p - 1$ and maximal $N_k$ reachable by my numerics $N_k <= 19$. As we learned in the previous sections, in the commensurate state there is one period of the lattice potential per one period of the spontaneous CDW. One can say that the incommensurate solution with 20 CDW's per 19 lattice periods would have exactly one excess period of spontaneous CDW structure per 19 unit cells as compared to the commensurate state on top of the same lattice. By inspecting this solution (Fig.\ref{fig:BulkProfs}, top) we see, that the solution profile coincides with the commensurate state almost everywhere except from the finite size region in the core, where this excess of 1 period of the spontaneous structure is accounted for. 

\begin{figure}[ht]
\includegraphics[width=0.49 \linewidth]{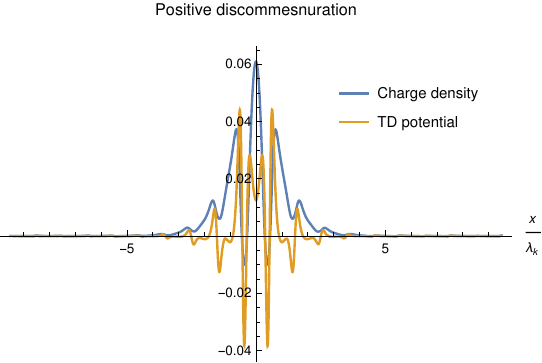}
\includegraphics[width=0.49 \linewidth]{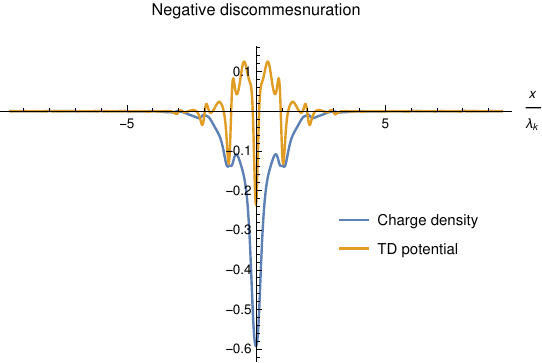}
\caption{\label{fig:Density}\textbf{Charge and thermodynamic potential of a discommensuration} measured with respect to the parent commensurate state. The distributions clearly show that the discommensuration is a local object. The positive disc. is considered at $A=1$, the negative -- at $A=2$. The charge is manifestly positive for positive disc. and manifestly negative for negative one, in analogy with the toy model.}
\end{figure}

\begin{figure}[ht]
\includegraphics[width=0.49 \linewidth]{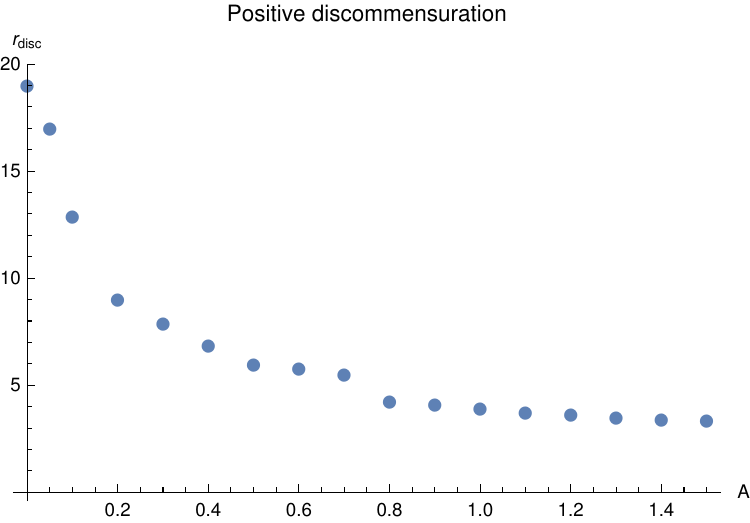}
\includegraphics[width=0.49 \linewidth]{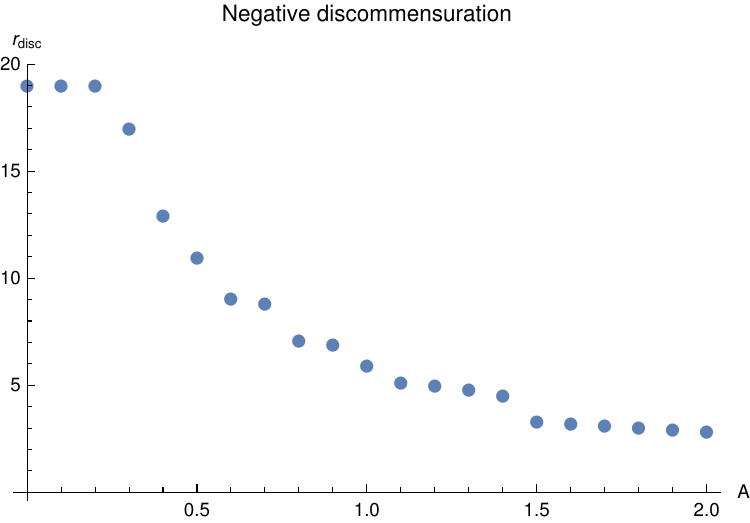}
\caption{\label{fig:DiscSize} \textbf{Size of the discommensuration}, obtained as a width at 0.1 height of the charge density (see Fig.\ref{fig:Density}), depending on the amplitude of the lattice. As the lattice becomes stronger, the discommensuration localizes, in complete analogy with the toy model.}
\end{figure}

We can also study the TD potential and charge density of such solutions as compared to the pure commensurate ``parent'' state. Fig.\ref{fig:Density} shows clearly that this incommensurate solution can be seen as a commensurate state with one localized \textit{discommensuration}(disc.) on top of it. Similarly, the solution with 18 CDW's per 19 unit cells includes a single discommensuration with deficiency of 1 CDW. As seen on Fig.\ref{fig:DiscSize}, the size of a disc. does indeed decrease when the lattice gets stronger as anticipated from our toy model. 

From this point of view the discommensuration is a soliton on the commensurate background with a topological charge $\pm 1$, coinciding with the number of missing/excess periods of the spontaneous charge modulation in the domain. We observe both of these types of discommensurations in our model. The positive disc. appears when one considers wavelengths of the ionic lattice, larger then the wavelength of the spontaneous crystal $\lambda_k > \lambda_{p_0}$, and the preferred commensurate fractions $N_p/N_k$ are larger then 1. The negative discommensuration is seen when $N_p/N_k<1$. Both types are direct analogues of discommensuration studied in the context of charge density waves in \cite{mcmillan1976theory}. 

\begin{figure}[ht]
\includegraphics[width=0.49 \linewidth]{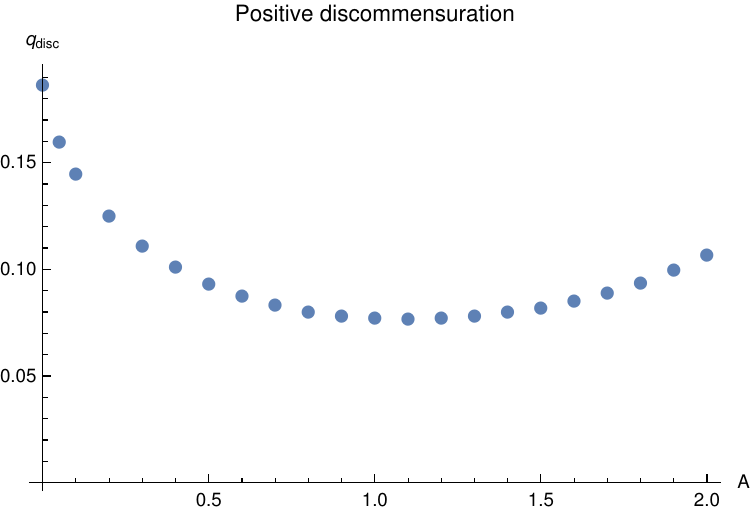}
\includegraphics[width=0.49 \linewidth]{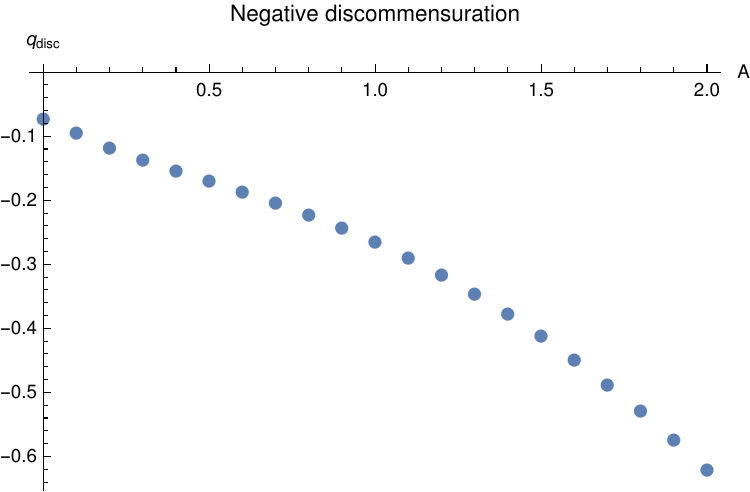}
\caption{\label{fig:Charge} \textbf{Charge of a single discommensuration} measured with respect to the ``parent'' commensurate state, depending on the amplitude of the lattice. There are no preferred values, as opposed to the naive expectations.}
\end{figure}

\begin{figure}[ht]
\includegraphics[width=0.49 \linewidth]{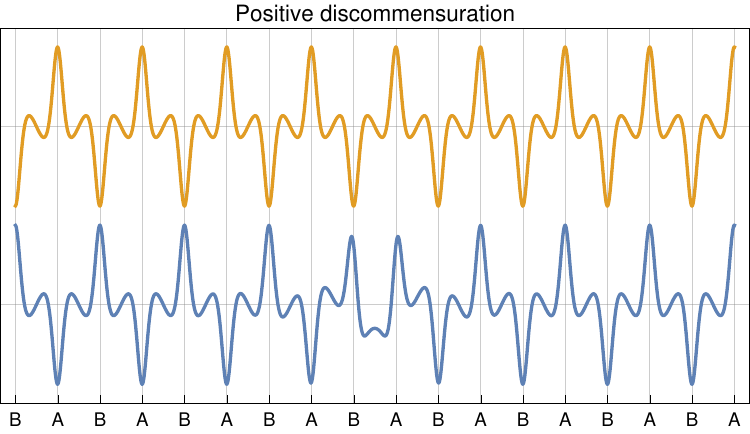}
\includegraphics[width=0.49 \linewidth]{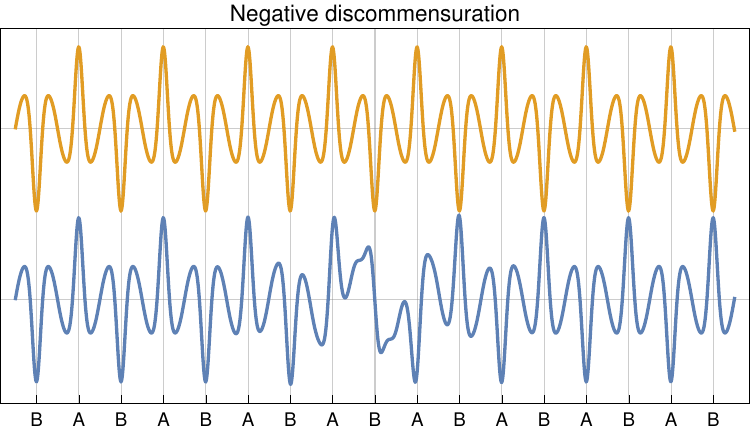}
\caption{\label{fig:Currents} \textbf{Current profile a discommensuration} (blue curves) as compared to the current profile of the ``parent'' commensurate state (yellow curves). Both cases clearly show a domain wall, which shift the staggered $A-B$ order by half of a period. The positive disc. is considered at $A=1$, the negative -- at $A=2$.}
\end{figure}

From Fig.\ref{fig:Density} it is apparent that the discommensuration carry a manifestly positive (negative) electric charge. The important difference between the holographic discommensuration and the trivial example discusse din the Introduction is that its charge \textit{is not fixed} and varies smoothly, see Fig.\ref{fig:Charge}. To complete the comparison of the holographic discommensuration with the toy model expectations, let's consider the current profile of these solutions, Fig.\ref{fig:Currents}. Clearly, disc. serves as a domain wall in the staggered current order.

At this moment we can understand the absence of one shoulder in the $\Omega(\lambda_p)$ curves, observed in Sec.\,\ref{sec:nonlin_solution}. Indeed, starting from the commensurate point one can move to higher stripe wave vectors by adding charge $+1$ disc., or to lower wave vectors by adding charge $-1$ disc. For any given set of the background parameters only one of them is dynamically stable. This can be understood from the simple energy balance argument: for $k > p_0$ the CDW in commensurate state has higher wave vector then the spontaneous one, hence lowering the wave vector by negative charge disc. would lower the potential energy. From the other hand, the shape of the disc. as a localized object contributes to kinetic energy. The balance between these contributions would stabilize the soliton with charge $-1$. If now one would consider the positive charge disc. on the same background, one would see that it rises the stripe wave vector, hence the potential energy is also rising, while the contribution from the kinetic energy is always positive. Hence there is no way the different energy contributions can be balanced in the soliton and it is not stable.

\begin{figure}[ht]
\centering
\includegraphics[width=0.8 \linewidth]{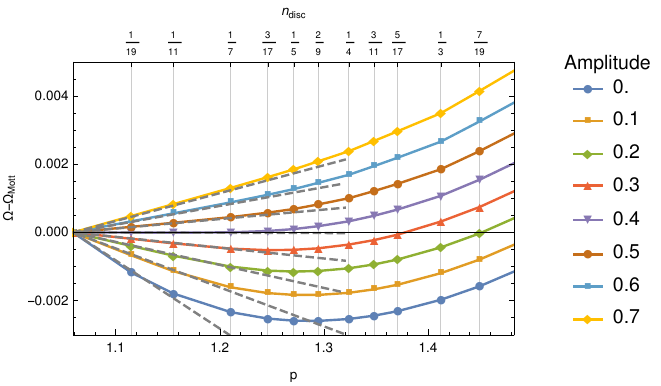}
\includegraphics[width=0.8 \linewidth]{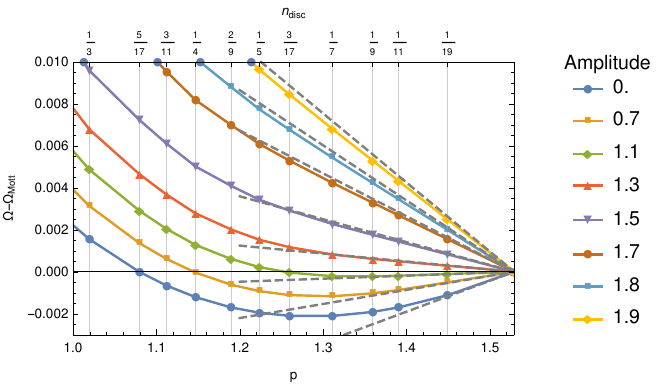}
\caption{\label{fig:wslopes} \textbf{Thermodynamic potential} as a function of the density of discommensurations, measured with respect to the commensurate state at different amplitudes of the lattice. For temperature $T=0.01$ the spontaneous momentum is $p_0(T) = 1.28$. Upper panel shows the case when the lattice momentum is smaller: $k =1.06<p_0$, and positive discommensurations increase the total momentum of the incommensurate solution.  Lower panel -- the lattice momentum is larger $k =1.53>p_0$ and negative discommensurations reduce the momentum of incommensurate solution. The slopes of the dashed lines correspond to the mass of a single discommensuraton. When this slope becomes negative, the commensurate state is unstable and the discommensurations proliferate.}
\end{figure}

\begin{figure}[ht]
\center
\includegraphics[width=0.49 \linewidth]{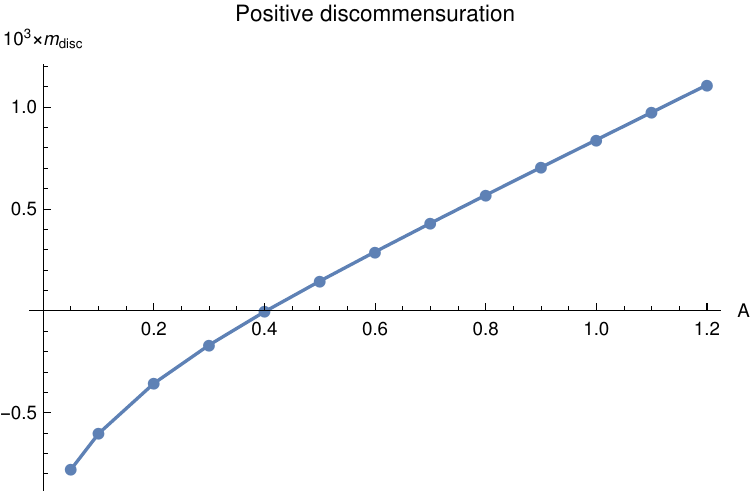}
\includegraphics[width=0.49 \linewidth]{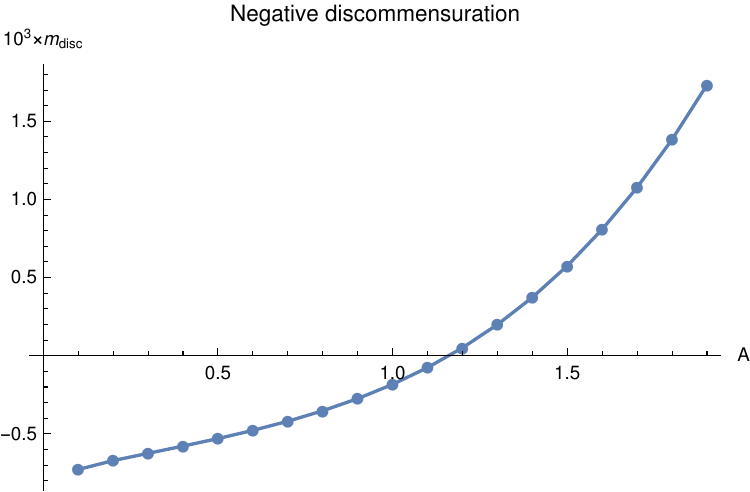}
\caption{\label{fig:mDisc} \textbf{Mass of an isolated discommensuration}, depending on the amplitude of the lattice. At small amplitude the mass is negative signaling that the commensurate state is unstable. At large amplitude discommensurations become massive and the pure commensurate state stabilizes.} 
\end{figure}

One can see that the further deviation from the commensurate point $1/1$, according to \eqref{discom_counting}, is achieved by rising the density of discommensurations (disc.) of the particular type, i.e. considering one disc. per fewer lattice periods. In this regard we can re-analyze the data which we obtained in Sec.\,\ref{sec:nonlin_solution}. Figure \ref{fig:wslopes} shows the thermodynamic potential of the state, measured w.r.t. the thermodynamic potential of the commensurate state, as a function of the density $n$ of disc.. While the density is low enough, and the separation between the solitons is much larger then their size, the apparent linear dependence of $\Omega(n)$ just follows from the fact that the isolated soliton has a certain fixed ``mass'', which is given by the slope of the $\Omega(n)$ curve at $n\rar 0$. The field profiles in this regime look as the domains of commensurate solutions separated by the evenly spaced lattice of discommensurations. It is also evident that the ``mass'' of a soliton depends on the parameters of the solutions and can be negative as well as positive, see Fig \ref{fig:mDisc}. The positive mass would mean that the creation of disc. costs energy and the commensurate state is therefore energetically stable. The negative mass, on the contrary, signals the instability of the commensurate state. The solitons start to proliferate and their density rises, forming the increasingly dense discommensuration lattices and driving the state further from the commensurate value of $1/1$. As the density rises, the distance between the solitons becomes smaller and they start to interact. The repulsion between disc. limits the energetically preferred density from above and in this way the system assumes the stable incommensurate stripe wavelength.

\section{Conclusion}
In this note I studied in detail the new solitonic object-- discommensuration -- which appears in the holographic model as soon as one considers the strong interplay between explicit and spontaneous symmetry breaking.
I demonstrated that in many regards these holographic discommensurations conform the naive expectations obtained from the classical toy model: they are localized objects, which are responsible to the mismatch between the periods of the explicit and spontaneous structures; they carry charge and they realize the domain walls in the staggered order parameter. The important difference however is the absence of the preferred value of charge, which would be discrete in the classical model. 

Discommensurations play an important role in the commensurate/incommensurate phase transition. The commensurate state is stable as long as the mass of disc. is positive. Once the strength of the lattice is lowered, the mass goes negative and discommensurations proliferate, forming the incommensurate state. 

The involved numerical analysis is needed to study discommensurations. Given that they are solitons on top of the nonlinearly constructed ground state, it would be interesting to find out whether any analytic control over them is possible. The examples of the analytical treatment of the similar solutions include \cite{Arancibia:2015saa,Arancibia:2014vua}. Also, to some extend they are similar to the Abrikosov vortices in the superconducting condensate. The latter can be analyzed perturbatively near the critical temperature, when the condensate itself is small. It would be interesting to consider the similar approach to discommensurations.

\acknowledgments

I'm grateful to Tomas Andrade, with whom we are constantly collaborating on various commensurability projects. I'd like to thank Koenraad Schalm, Jan Zaanen, Aristos Donos, Christiana Pantelidou and Niko Jokela for insightful comments and useful discussions. I appreciate hospitality of the HEP group in NORDITA, Stockholm and the organizers of the workshop 
``Many-Body Quantum Chaos, Bad Metals and Holography'', where the preliminary results of this study have been reported.

This work is supported in part by the VICI grant of Koenraad Schalm from the Netherlands Organization for Scientific Research (NWO), by the Netherlands Organization for Scientific Research/Ministry of Science and Education (NWO/OCW) and by the Foundation for Research into Fundamental Matter (FOM) and partially by RFBR grant 15-02-02092a. 

The numerical calculations have been performed on the Maris Cluster of Lorents Insititute.

\appendix

\section{Numerical techniques} 
\label{sec:numerical_techniques}

The present study relies heavily on numerical analysis of the holographic nonlinear solutions. Moreover, in order to study the phase diagram and cover the parameter space several thousands of the solutions to equations of motion have been obtained, with some of them requiring quite large calculation grids in the spatial direction. This situation puts a very strict requirements on the numerical techniques which are used and the precision and accuracy of the results. In this Appendix I review the key features of the numerical setup, used in the present study.

Roughly speaking, the process of the numerical solution of the system of nonlinear PDEs consists of a few key steps \cite{boyd2001chebyshev}:
\begin{enumerate}
	\item \label{num:discretize}Given an approximation to the solution (in the form of the values of functions at the grid nodes), evaluate the derivatives of the functions. This step requires choosing the appropriate discretization scheme. 
	\item \label{num:eval_coef} Given the values of functions and derivatives at the nodes, calculate the coefficients in the linearized equations and the values of the full nonlinear equations. For nonlinear problem one has to evaluate these coefficients anew at every step. This operation requires as many evaluations as the number of nodes and thus deserves optimization.
	\item \label{num:inverse} At this moment the problem can be written as system of linear algebraic equations. One can solve it either exactly by computing the inverse of the linear operator matrix (this corresponds to the Newton-Raphson method), or approximately, using some kind of "relaxation" scheme: the various options include (preconditioned) Richardson relaxation, Gauss-Seidel iterations or ILU decomposition.
	\item \label{num:step} Once the inverse is computed, the increment in the functional variables can be evaluated, which is used to construct the next approximation to the solution.
	\item \label{num:control} The iterations continue until some criterion of the accuracy or precision of the current approximation is fulfilled.
 \end{enumerate} 
In order to build an efficient solver one needs to choose carefully the tactics at every step.

Starting with the discretization scheme of step \ref{num:discretize} one usually chooses between pseudospectral collocation \cite{trefethen2000spectral} and FDD (finite difference derivative) of a certain order. The advantage of pseudospectral scheme is its improved accuracy, hence one needs much less grid points in order to approximate the solution well. In some cases this method also demonstrates the exponential convergence. Pseudospectral discretization works perfectly in the periodic spatial direction, where the solution is smooth and is well approximated by a Fourier series. The drawback is the efficiency of the scheme on the finite interval of the holographic coordinate. Here one has to use the smooth Chebyshev polynomials, which are not suitable for approximation of the non-analytic behavior near the UV boundary and IR horizon. This mismatch thwarts the exponential convergence and may lead to the breakdown of the whole scheme. The FDD approach doesn't suffer from this drawback but requires much more grid points to reach the comparable accuracy. This has undesirable side effects on steps \ref{num:eval_coef} and \ref{num:inverse}, when the number of evaluations of coefficients and the size of the linear matrix to be inversed are increased, correspondingly. It is worth mentioning though, that FDD scheme leads to much sparser linear matrix on the step \ref{num:inverse}. If one chooses to inverse the matrix exactly, this task is next to impossible for sizable grids in the pseudospectral case, where the differentiation matrices are dense. The compromise would be to use several patches along the holographic direction \cite{Donos:2014yya}, using pseudospectral approximation in the interior and FDD near the boundaries. After experimenting with all three options, see below, I've chosen a single patch pseudospectral scheme in the holographic direction, which proved to be quite robust in practice.

I use \texttt{Wolfram Mathematica}\cite{Mathematica10} in order to implement the numerical algorithm. It may have the disadvantage in speed, when one compares it to the lower level computing languages like \texttt{C++} or \texttt{FORTRAN}, but as long as one uses the high level efficient precompiled routines like: \texttt{LinearSolve}, \texttt{NDSolve`FiniteDifferenceDerivative}, and sparse matrix operators the difference in speed becomes less obvious. The elementwise operations required at step\,\ref{num:eval_coef} can be efficiently compiled with \texttt{Compile}, which brings up a spectacular acceleration. In the end of the day, the most important limitation of Mathematica is the necessity to work with \texttt{MachinePrecision} numbers in the compiled function, which eventually limits the precision of the results, as discussed below. 

As we've discussed already, the direct inversion (Newton-Raphson method) in case of pseudospectral discretization is extremely demanding for the large grid which we use. Moreover, since we are solving the nonlinear problem, which inevitably requires an iterative procedure, obtaining very precise result for the matrix inverse at step\,\ref{num:inverse} doesn't make much sense. That's why we use a relaxation scheme instead. The speed of convergence of a relaxation scheme is defined by the highest eigenvalues of the linear operator matrix. In order to make the process more efficient one uses the preconditioning, i.e. one multiplies the operator by the preconditioner matrix, which brings all the eigenvalues to the same scale. The ideal preconditioner is the inverse of the operator itself, but a reasonable approximation to it will also work fine. I use the differential operator evaluated in the low order FDD scheme, as a preconditioner. It approximates the highest eigenvalues very well and is relatively easy to invert, being very sparse. The result is a nonlinear Richardson relaxation with Orszag preconditioning (See Sec. 15.14 and eq.(15.115) in \cite{boyd2001chebyshev}). One can also view this approach as a pseudo-Newton method, where the approximation to Jakobian is used instead of the exact one. It should be stressed here, that even though a low order FDD scheme is used in the construction of the linear operator, the equations to be solved use the full pseudospectral approximation to the derivatives, thence the pseudospectral accuracy is  achieved by iteratively inverting the sparse operator on a relatively coarse grid. The relaxation scheme requires, in principle, more iteration steps than the Newton-Raphson, so effectively the memory consumption is excanged with the CPU time. Altogether for the hardware which I used the relaxation procedure turned out to be an order of magnitude faster then the analogous Newton-Raphson scheme.

As mentioned already, the high eigenvalues of the linear operator are well approximated using the low level FDD preconditioner. This is not true for the lowest eigenvalues, which define the long-wavelength errors with slowest relaxation rate. One can fight these ones and further improve the efficiency of the numerical scheme by using multigrid technique \cite{briggs2000multigrid}. In practice I found that without multigrid the solutions converge already after $\sim 20$ steps. In this situation the overhead of transitioning between fine and coarse grids becomes relatively significant and I found no improvement of the overall efficiency in full multigrid method.

In the end of the day the calculation scheme was optimized to the extent when it takes about half an hour to obtain the precise solution on the largest grid of size $\sim 330_x \times 80_z$ (pseudospectral) using a single core of a laptop CPU (Intel Core i7-5600U @ 2.60GHz ) and about 3 Gb of RAM.


\section{Precision control} 
\label{sec:precision_control}
As one can see from the results, Fig.\,\ref{fig:wcurves}, the difference between the TD potentials of the solution with spontaneous structure and without it is just of order of few percents of the TD potentials themselves. That means that in order to reliably study this difference, the potentials must be evaluated with accuracy of at least $10^{-4}$. This puts a challenge to the numerical scheme and renders the strict precision control a necessity. 

The accuracy of the numerical results depends substantially on the size of the grid. On one hand, it is clear that the denser grid delivers higher accuracy, i.e. the closer approximation of the real result of continuum limit. On the other hand, very dense grids bring up numerical precision issues, when the rounding errors play a dominant role in the calculation of the derivatives. Thus the grid size must be optimized maximizing the accuracy while keeping the rounding errors below the certain value. This study also helps to choose between different methods of discretization in the holographic direction, discussed earlier.

In 2-dimensional problem the grid size has to be optimized in two directions: $N_x$ and $N_z$. As an example we use a sample solution describing the commensurate stripe on top of the lattice with amplitude $A=1.$, lattice wave-vector $k=1.5$ at temperature $T=0.01$. We perform a relaxation of this solution on the set of grids with $N_x = \{9, 17, 33, 65,129\}$ and $N_z = \{20,30,40,60,80\}$ (in pseudospectral case) or $N_z = \{40,80,120,160,240,320\}$ (in FDD case). For every calculation we keep track of the following convergence criteria: 
\begin{align*}
&\max |df|: & & \mbox{ Maximum value of the functional increment at every iteration step} \\
&\max |\sqrt{\xi^2}|:& & \mbox{ Local norm of the DeTurck vector} \\
&\max |G_\mu^\mu|:& &\mbox{ Local trace of the Einstein equation} \\
&d \Omega / \Omega :& &\mbox{ Relative increment of the mean thermodynamic potential at every step}
\end{align*}

\begin{figure}[ht]
\center
\includegraphics[width = 0.6 \linewidth]{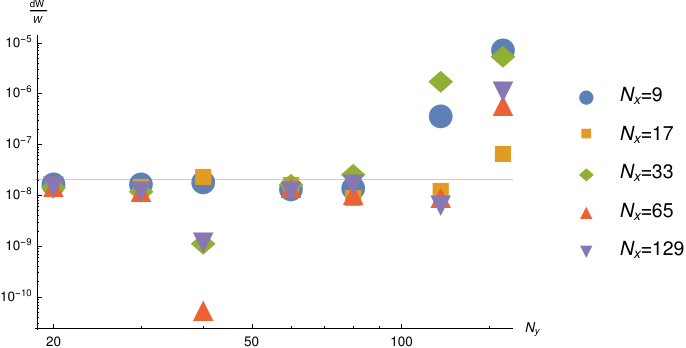}
\caption{\label{fig:round_err} Precision of the thermodynamic potential limited by the rounding errors for different Chebyshev grid sizes.}
\end{figure}

Each time the iteration procedure is run until $\max|df|$ hits the \texttt{MachinePrecision} bound $\sim 10^{-11}$, while the values of the functions are of order one. Due to the numerical rounding errors the precision of the solution can not be improved after this bound is reached. By studying the value of $d\Omega /\Omega$ at this point one can estimate the highest precision of the physical observables, which can be obtained on a given grid. On Fig.\,\ref{fig:round_err} the relative precision of the thermodynamic potential is shown for various grid sizes in pseudospectral approach. One can see that for dense grids $N_y > 80$, the relative error is increasing. No similar effect is seen for the dense $N_x$ grid. This important observation tells us that with given \texttt{MachinePrecision} there is no reason to use the grids with $N_y > 80$ and also that the numerical precision of the thermodynamical potential, which we calculate is not better then $10^{-8}$.

\begin{figure}[ht]
\center
\includegraphics[width = 0.48 \linewidth]{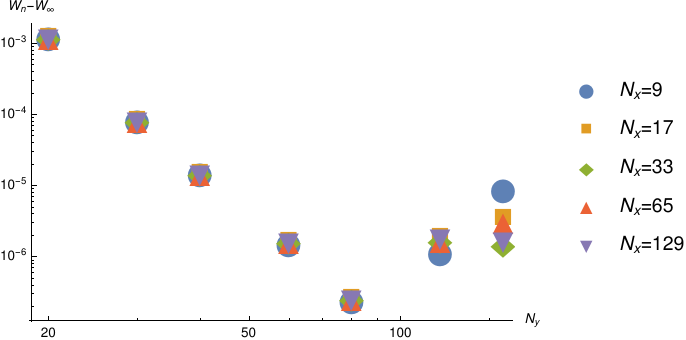}
\includegraphics[width = 0.48 \linewidth]{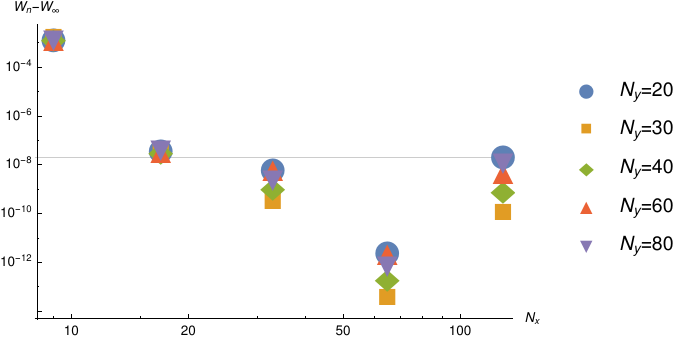}
\caption{\label{fig:accuracy} Accuracy of the thermodynamic potential depending on the size of grid in holographic ($N_y$) and spacial ($N_x$) direction.}
\end{figure}

In principle, the accuracy of the numerical results must increase with increasing grid size, giving the exact match in continuum limit. In order to estimate the exact result $\Omega_\infty$ we evaluate the thermodynamic potential for a set of increasing grid sizes and then extrapolate to infinity. The accuracy for a given grid is then defined by $|\Omega - \Omega_\infty|/\Omega_\infty$. As one can see from Fig.\,\ref{fig:accuracy}, the optimal accuracy reached at $N_y = 80$ is of order $10^{-7}$. As to the spatial grid resolution, one can see that already for $N_x=33$ the result is close enough to $\Omega_\infty$, i.e. already at $N_x = 33$ the accuracy is controlled by the holographic axis resolution $N_z$. 

In the end of the day it's clear that for a single patch Chebyshev grid the maximum $N_y$ resolution is limited by the rounding errors at $N_y=80$. The accuracy of the thermodynamical potential for a grid of this size is about $10^{-7}$. I use this value as a numerical error estimate throughout the present study and it proves to be quite enough for the main results. Also, the $x$-axis resolution can safely be set to $N_x=33$ points per one period of the CDW, without affecting the accuracy of result. It should be noted here that the comparable accuracy in FDD approach is reached for $N_y \approx 320$. Using the grid of this large size is disadvantageous on the other stages of calculation, therefore I rely on the pseudospectral approach instead. I've also checked that the patching technique doesn't bring any significant improvement of the accuracy.

One should keep in mind that in the numerical procedure the DeTurck equations are solved, so it must be checked that the Einstein equations are satisfied. This is done by the two independent measures:  $\max |G_\mu^\mu|$ and $\max |\sqrt{\xi^2}|$. For not very low temperatures $T> 0.01$ at the grids which were used these values are both of order $10^{-7}$, which is quite satisfactory. However, at lower temperatures they increase and the use of higher $N_y$-resolution grid is necessary. As stated above, the $N_y$-resolution is bounded by the rounding errors and \texttt{MachinePrecision}, henceforth with the present numerical scheme I can not reliably access extremely low temperatures. Nonetheless, this drawback has little impact on the results of the present study.


\bibliographystyle{JHEP-2}
\bibliography{inhom_stripes_lattice}

\end{document}